\documentclass[titlepage,12pt]{article}

\usepackage{authblk}
\usepackage[margin=1in]{geometry}

\usepackage{amsbsy,amsfonts,amsmath,amssymb,amsthm}
\usepackage{bm}
\usepackage[inline]{enumitem}
\usepackage[colorlinks=true,citecolor=blue]{hyperref}
\usepackage{soul}
\usepackage{xcolor}
\usepackage{setspace}

\usepackage{dsfont}

\usepackage{booktabs,threeparttable}
\usepackage{multirow}
\usepackage{graphicx}
\graphicspath{{./}, {./image/}}
\usepackage{tikz}

\usepackage{natbib}

\usepackage{color}

\usepackage[pagewise]{lineno}
\linenumbers*[1]
\newcommand*\patchAmsMathEnvironmentForLineno[1]{%
	\expandafter\let\csname old#1\expandafter\endcsname\csname 
	#1\endcsname
	\expandafter\let\csname oldend#1\expandafter\endcsname\csname 
	end#1\endcsname
	\renewenvironment{#1}%
	{\linenomath\csname old#1\endcsname}%
	{\csname oldend#1\endcsname\endlinenomath}}%
\newcommand*\patchBothAmsMathEnvironmentsForLineno[1]{%
	\patchAmsMathEnvironmentForLineno{#1}%
	\patchAmsMathEnvironmentForLineno{#1*}}%
\AtBeginDocument{%
	\patchBothAmsMathEnvironmentsForLineno{equation}%
	\patchBothAmsMathEnvironmentsForLineno{align}%
	\patchBothAmsMathEnvironmentsForLineno{flalign}%
	\patchBothAmsMathEnvironmentsForLineno{alignat}%
	\patchBothAmsMathEnvironmentsForLineno{gather}%
	\patchBothAmsMathEnvironmentsForLineno{multline}%
}

\nolinenumbers

\allowdisplaybreaks 

\newcommand{\pkg}[1]{{\normalfont\fontseries{b}\selectfont #1}}
\let\proglang=\textsf

\begin{document}
	
\title{Comparison of sectoral structures between China and Japan: A network
  perspective}

\author[1]{Tao Wang}
\author[2,$\ast$]{Shiying Xiao}
\author[2]{Jun Yan}
\affil[1]{School of Statistics, Shanxi University of Finance and
  Economics, Taiyuan 030006, China}
\affil[2]{Department of Statistics, University of Connecticut, Storrs, CT
  06269, USA}
\affil[$\ast$]{Corresponding author. Email:
\href{mailto:shiying.xiao@uconn.edu}{shiying.xiao@uconn.edu}}

\maketitle
	
\doublespacing
	
\begin{abstract}
Economic structure comparisons between China and Japan have long captivated
development economists. To delve deeper into their sectoral differences from
1995 to 2018, we used the annual input-output tables (IOTs) of both
nations to construct weighted and directed input-output networks (IONs). This
facilitated deeper network analyses. Strength distributions underscored
variations in inter-sector economic interactions. Weighted, directed
assortativity coefficients encapsulated the homophily among connecting sectors'
features. By adjusting emphasis in PageRank centrality, key sectors were
identified. Community detection revealed their clustering tendencies among the
sectors. As anticipated, the analysis pinpointed manufacturing as China's
central sector, while Japan favored services. Yet, at a finer level of the specific
sectors, both nations exhibited varied structural evolutions. Contrastingly,
sectoral communities in both China and Japan demonstrated commendable stability
over the examined duration.

\end{abstract}

\section{Introduction}
\label{sec:intro}

China and Japan are both major players in the global economy. While China has a
larger economic scale and faster growth rate, Japan has a more advanced
industrial structure and higher labor productivity~\citep{chansarn2010labor}. As
of 2022, China's Gross Domestic Product (GDP) in current prices was 18.32
trillion US dollars (USD), with 12,970 USD per capita; Japan's GDP was 4.3
trillion USD, with 34,358 USD per capita~\citep{IMF2023gdp}. The service sectors
hold a larger share in Japan's economy at 69.5\%, compared to China's 54.5\%,
indicating that Japan's sectoral structure is more advanced 
\citep{chenery1986industrialization}. Although China and Japan follow distinct
development paths and possess different natural resources, there are intriguing
parallels between China's recent economic growth and Japan's historical
development, suggesting that potential lessons from Japan's experience could
benefit China~\citep{minami2010thelewis, fukumoto2012rebalancing}. 
For example, the move from a manufacturing-based economy to a 
knowledge- and service-based economy was vital for Japan to maintain its
economic momentum~\citep{alvstam2009economic}; the structural problems due to
the aging demographic in Japan led to sluggish economic growth and recession
since the early 1990s~\citep{yoshino2016causes}. Many insights and lessons from
Japan could be learned by China to prolong its economic growth and prevent
long-term recessions through comparing the economic structures of China and
Japan over time~\citep[e.g.,][]{fujita2004spatial, ogunmakinde2019review}.

While there is a plethora of research focused on the economies of China and
Japan, comparative studies on their sectoral structures remain sparse. The
existing literature extensively covers various facets of both economies,
encompassing trade relationships~\citep{howe1990china,
  marukawa2012bilateral,katz2013mutual}, economic influences in other
regions~\citep{dreger2014does, wang2022china}, investment
strategies~\citep{fung2002determinants, katada2020china}, and financial
markets~\citep{okimoto2009financial, schnabl2017exchange}. However, direct
juxtapositions of their sectoral structures are notably limited. Of the few
studies that exist, the majority lean on traditional input-output table (IOT)
methodologies~\citep{leontief1963structure, miernyk1965elements}. For instance,
\citet{min2019comparative} delved into the industrial spillover effects between
the two nations, while \citet{liu2018method} spotlighted their core industrial
structures. Surprisingly, network analysis tools, which are inherently suited
for IOT analyses, have been underutilized in this comparative context. A rare
exception is \citet{li2017examining}, but their scope was confined to
quantitatively probing evolutionary trends in industrial structures.

A modern and powerful tool to analyze IOTs is network
analysis~\citep[e.g.,][]{newman2003structure, boccaletti2006complex} since
an IOT naturally defines an input-output network (ION). 
By converting the IOT's sectors into nodes and
transaction flows into edges, it facilitates the creation of an economic network
graph, thereby illuminating the intricate relationships and dependencies
intrinsic to the ION~\citep{schweitzer2009economic, contreras2014propagation,
  xu2019input, wang2021regional, li2022manufacturing}. The salient advantages of
this approach include its ability to vividly present data, translating
input-output information into intuitive and comprehensible graphical
forms~\citep{cruz2014community}. Moreover, it offers a deeper understanding of
sectoral interdependencies, extending beyond the simple metrics of traditional
input-output analysis. This depth encompasses global metrics, such as directed
strength distributions and weighted, directed assortativities, and
sector-specific centrality measures. Finally, network analysis introduces novel
techniques for pinpointing key sectors and discerning sectoral clusters, 
techniques
that eclipse traditional input-output methodologies~\citep{muniz2008key,
  xiao2022incorporating}. The key sector identification further allows for a
tailored approach where key sector identification can be adapted based on
specific analytical objectives in centrality metrics~\citep{zhang2022pagerank,
  xiao2022incorporating}. Embracing all these network analytic tools provides an
unparalleled framework for comparing the sectoral architectures of nations
based on their IONs.

This paper compares the sectoral structures between China and Japan, leveraging
both traditional and newly developed network analysis methodologies. While
existing research in this domain has
primarily focused on degree distribution and key sectors, these studies
often overlook the directionality and weight of IONs. Moreover, they tend to
bypass auxiliary information when identifying key sectors and largely omit
sectoral clustering within IONs. To address these gaps, we harness IOTs
from China and Japan spanning 1995--2018, presenting three major
contributions. First, we juxtapose the characteristics of inter-sectoral
connections in both nations, focusing on node strength distribution and the
cutting-edge concept of weighted, directed assortativity
coefficients~\citep{yuan2021assortativity}, along with jackknife standard
deviations~\citep{lin2020theoretical}. Second, we identify the top five key
sectors and trace their evolution in both countries, drawing upon the recently
introduced PageRank (PR) centrality measure~\citep{zhang2022pagerank}.
Lastly, we group tightly-knit sectors into communities using an optimal
modularity algorithm tailored for weighted, directed
networks~\citep{newman2006modularity}, and assess the similarities of the
community structures over time in each nation. Together, our analyses furnish a
comprehensive understanding of China and Japan's sectoral dynamics, elucidating
their distinctions from a network vantage.

The paper is structured as follows. Section~\ref{sec:ion} introduces the IOT
data source and the resulting IONs. In Section~\ref{sec:strength}, we compare
the strength distributions of sectors between China and
Japan. Section~\ref{sec:assort} investigates the assortative characteristics of
sector connections while taking uncertainties into account with a jackknife
approach. Section~\ref{sec:key} ranks important sectors in each country by a
PR centrality measure that incorporates auxiliary
information. Section~\ref{sec:community} compares clusterings of sectors through
community detection. Finally, Section~\ref{sec:dis} concludes with a discussion.

\section{ION data}
\label{sec:ion}

An IOT is a valuable tool for analyzing the interdependent relationships and
structures between different sectors within an economy by tracking the monetary
transactions between them. The basic structure of an IOT for an economy with $n$
sectors is shown in Table~\ref{tab:niot}. The IOT consists of three parts:
(1) the intermediate use matrix $W:= (w_{ij})_{n \times n}$,
where $w_{ij}$ denotes the cost of the products or services that sector~$i$
provides to sector~$j$, $i, j \in \{1, \ldots, n\}$, and each row~$i$ and
column~$j$ contain the amount of value that sector~$i$ provides to and consumes
from other sectors, respectively;
(2) the final use $F:=(f_i)_{n \times 1}$, which is the horizontal extension
of~$W$, with each $f_i$ represents the product produced by sector~$i$ for
consumption, investment, and net export; and
(3) the value added $X^\top:=(x_j)^\top_{n \times 1}$, which is the vertical
extension of~$W$, with each $x_j$ represents the value-added by sector~$j$.
The total output, which equals the total input, is represented by
$Y := (y_i)_{n \times 1}$. The table satisfies both row and column balance, with
the former stating that total output equals the sum of intermediate use and
final use for each sector, and the latter indicating that total output equals
the sum of intermediate use and value added for each sector.

\begin{table}[tbp]
\renewcommand*{\arraystretch}{1.5}
\centering
\caption{Fundamental structure of a national IOT.}
\label{tab:niot}
        \medskip
\begin{tabular}{c|c|c|c|c|c|c}
\hline
\multicolumn{2}{c|}{\multirow{2}{*}{}}
& \multicolumn{3}{c|}{Intermediate use}
& \multirow{2}{*}{Final use}
& \multirow{2}{*}{Total output} \\
\cline{3-5}
\multicolumn{2}{c|}{} & Sector 1 & $\cdots$ & Sector $n$ & & \\
\hline
\multirow{3}{*}{Intermediate input}
& Sector 1
& \multicolumn{3}{c|}{\multirow{3}{*}{$W$}}
& {\multirow{3}{*}{$F$}}
& {\multirow{3}{*}{$Y$}} \\
\cline{2-2}
& $\vdots$ & \multicolumn{3}{c|}{} & &  \\
\cline{2-2}
& Sector $n$ & \multicolumn{3}{c|}{} & &  \\
\hline
\multicolumn{2}{c|}{Value added}
& \multicolumn{3}{c|}{$X^{\top}$}
& \multicolumn{2}{c}{} \\
\cline{1-5}
\multicolumn{2}{c|}{Total input}
& \multicolumn{3}{c|}{$Y^{\top}$}
& \multicolumn{2}{c}{} \\
\hline
\end{tabular}
\end{table}

The intermediate use matrix $W$ of an IOT defines a weighted, directed
ION $G(V, E, W)$, where each node $v_i\in V$ represents a sector and each edge
$e_{ij}\in E$ represents a transaction from the source sector $i$ to the target
sector $j$ with a weight of $w_{ij}$. The edge structure of the ION is described
by the adjacency matrix $A :=(a_{ij})_{n \times n}$, where $a_{ij}=1$ if
$e_{ij}\in E$ and $a_{ij}=0$ otherwise. Analyses of the IONs determined by the
IOTs provide a network perspective in studying the sectoral transaction
structures of economies, and can provide insights into topics such as
propagation of economic shocks and the role of sectors in the overall economy
\citep[e.g.,][]{contreras2014propagation}.

To examine the sectoral structures of China and Japan, we must construct IONs
from consistent IOTs. However, direct comparisons using each country's IOTs are
problematic: China updates its IOT every five years, the most recent one for
2017~\citep{NBS2023csy}, whereas Japan's data is available for 2005, 2011, and
2015~\citep{MIC2019iot}. Additionally, the unique economic landscapes and
sectoral details in each country mean their IOTs feature different sector
configurations. International IOT databases, like the World Input-Output
Database (WIOD)~\citep{timmer2015illustrated} and the Asian Development Bank's
Multiregional Input-Output database (ADB-MRIO)~\citep{ADB2023iot}, offer
alternative sources. The diagonal blocks of these international IOTs can be used
to construct individual national IOTs. However, these too have limitations; for
instance, the WIOD only spans 2000--2014; and the ADB-MRIO database covers 35
sectors, while China's data includes only 33 of these sectors.


The IOTs that we used are from the STructural ANalysis (STAN) database, a
valuable resource for IOTs from many countries~\citep{OECD2021stand}. Based on
the International Standard Industrial Classification of all economic activities,
Revision 4 (ISIC Rev.4)~\citep{UN2008isic}, the current version of the STAN
database includes annual IOTs with 45 sectors in a unified form for member
countries. The STAN database was constructed using annual national accounts by
activity tables from member countries, and other data sources such as national
industrial surveys and censuses were used to estimate missing
quantities~\citep{antras2012measuring}. Transaction amounts are reported in USD
of 2015, and data is available from 1995 to 2018.
The data from China and Japan are publicly available in the \proglang{R} package
\pkg{ionet}~\citep{pkg:ionet}.
However, it should be noted
that the 45th sector, ``activities of households as employers; undifferentiated
goods- and services-producing activities of households for own use'', is missing
for both China and Japan, resulting in 44 sectors for their respective IONs. A
description of the 44 sectors used in the analyses for China and Japan from 1995
to 2018 is provided in Table~\ref{tab:stan} in the Appendix~\ref{sec:seccode}. 
Based on the STAN database, we constructed IONs of both countries including 44
sectors form 1995 to 2018.

\section{Node strength analysis}
\label{sec:strength}

We begin by examining the differences in the node strength distribution. For a
directed and weighted network, the strength of a node $i$ can be partitioned
into its in-strength and out-strength, which represent the total weight of
incoming and outgoing edges, respectively. Specifically, we define
$s_i^{\text{in}}$ and $s_i^{\text{out}}$ as the sums of weights of all edges that
are incoming to and outgoing from node $i$, respectively. In the context of
IONs, $s_i^{\text{in}}$ and $s_i^{\text{out}}$ correspond to the total monetary
input and output, respectively, that sector~$i$ receives from and supplies to
other sectors. We can also define the total strength of a node as
$s_i=s_i^{\text{in}}+s_i^{\text{out}}$.

\begin{figure}[tbp]
\centering
\includegraphics[width=\textwidth]{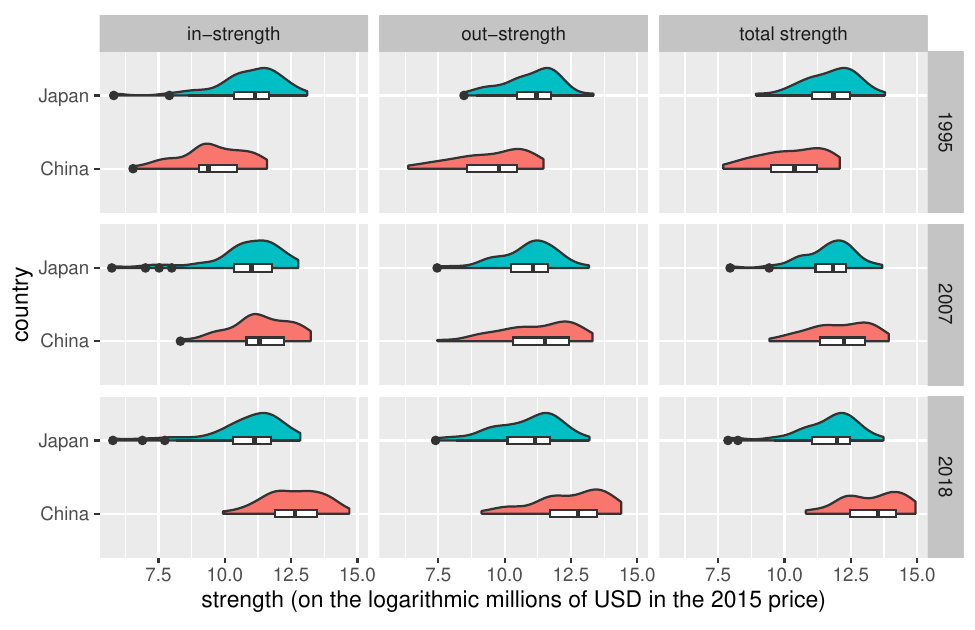}
\caption{Half-violin plots with boxplots overlaid, for in-, out-
	and total-strength (in million USD of the 2015 price on the log scale)
	of the IONs of China and Japan in 1995, 2007 and 2018.}
\label{fig:str}
\end{figure}

Figure~\ref{fig:str} displays half-violin plots~\citep{pkg:gghalves}
of the in-, out-, and total-strength
for 44 sectors of China and Japan in 1995, 2007, and 2018, measured in millions
of USD in 2015 price and using a log scale. In 1995, the distributions of all
three strengths overlapped, but with Japan's distribution clearly shifted to the
right, indicating larger monetary linkages between sectors in Japan compared to
China. By 2007, the two countries' distributions largely overlapped, but with
China's distributions now slightly to the right of Japan's. By 2018, the
overlapping areas decreased significantly, and China's distributions were
considerably shifted to the right of Japan's, suggesting that between-sector
economic flows in China had surpassed those in Japan. Vertical comparisons
across years reveal little change in the strength distribution for Japan, but
notable increases for China. These increases reflect the rapid growth of China's
economy, leading to expansion in the sizes of its sectors and between-sector
connections.

\begin{figure}[tbp]
\centering
\includegraphics[width=\textwidth]{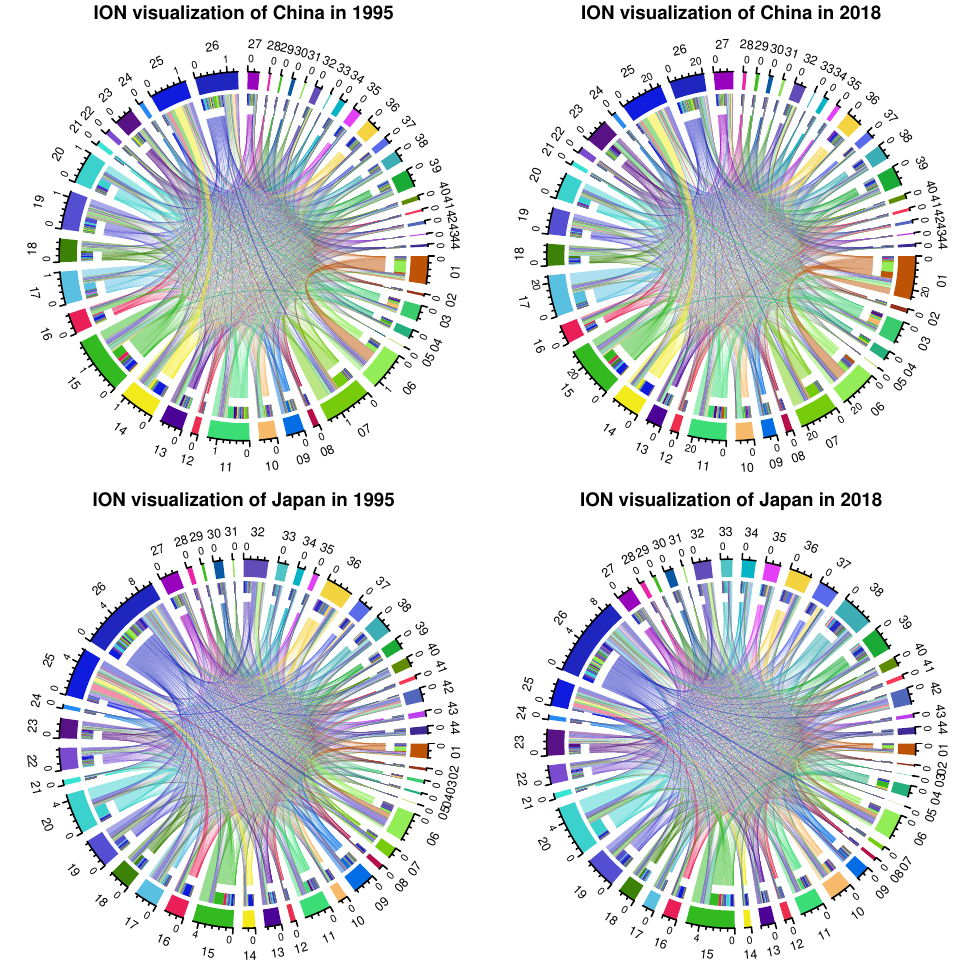}
\caption{ION visualization of China and Japan in 1995 and 2018.
The width of the chord connecting the arcs is directly proportional to
the magnitude of economic flow. Longer arcs represent greater outputs.
The unit of economic flow is 100 billion USD of the 2015 price.}
\label{fig:chord}
\end{figure}

Chord plots~\citep{gu2014circlize},
as seen in Figure~\ref{fig:chord}, effectively visualize the
nuanced inter-sectoral connections in the IONs of both China and Japan for 1995
and 2018. Each uniquely colored outer arc corresponds to a sector, with its
width indicating the sector's total strength. Connecting chords symbolize
inter-sectoral flow, their bandwidths being proportionate to the strengths,
and their colors matching the source sectors. A few observations stand out.
In both years, manufacturing holds a larger chunk of China's total strength than
Japan's. Conversely, Japan's service industry dominates in total strength over
China's. Sectorally, China's ``agriculture, hunting, forestry''~(01) claims a 
more
significant share than in Japan, a share which grew by 2018. Meanwhile, Japan's
``wholesale and retail trade; repair of motor vehicles''~(26) takes the lead.
Examining supply, or sectoral out-strength, China's primary contributors include
``basic metals''~(15), ``agriculture, hunting, forestry''~(01), ``wholesale
and retail trade; repair of motor vehicles''~(26), and ``chemical and chemical
products"~(11). For Japan, leading sectors are ``wholesale and retail trade;
repair of motor vehicles"~(26) and ``basic metals''~(15).
From a demand perspective (in-strength), ``construction''~(25), ``food products,
beverages and tobacco''~(06),  and ``basic metals''~(15) are predominant in
China, whereas in Japan, ``wholesale and retail trade; repair of motor
vehicles''~(26), ``motor vehicles, trailers and semi-trailers''~(20), and
``construction''~(25) lead. In general, sectoral linkages in two countries have
distinct characteristics, necessitating detailed network analyses for a
comprehensive quantitative assessment.

\section{Assortative mixing properties}
\label{sec:assort}

The assortativity of a network measures the homophily of a network, that is, the
tendency of nodes to connect with similar partners in a network.
A commonly used assortativity measure is the degree-degree
correlation~\citep{newman2002assortative}, which is easily adapted to total
strength assortativity by replacing the degrees with strengths.
Since IONs are directed and weighted, we further use a class of weighted,
directed assortativity measures calculated with node strengths
\citep{yuan2021assortativity}:
\begin{equation*}
r_{\alpha, \beta} = 
\frac{\sum_{i,j \in V} w_{ij}
\left(s_i^{(\alpha)} - \bar{s}_{\rm sou}^{(\alpha)}\right)
\left(s_j^{(\beta)} - \bar{s}_{\rm tar}^{(\beta)}\right)}
{\sqrt{\sum_{i,k \in V} w_{ik}
\left(s_i^{(\alpha)} - \bar{s}_{\rm sou}^{(\alpha)}\right)^2}
\sqrt{\sum_{k,j \in V} w_{kj}
\left(s_j^{(\beta)} - \bar{s}_{\rm tar}^{(\beta)}\right)^2}
},
\end{equation*}
where $s_i$ and $s_j$ is the pair strengths of nodes $i$ and $j$ corresponding 
to edge $e_{ij}$,  $(\alpha,\beta) \in \{\text{in}, \text{out}\}$ 
index strength type, and
\begin{equation*}
\bar{s}_{\rm sou}^{(\alpha)} = 
\frac{\sum_{i,j \in V} w_{ij} s_i^{(\alpha)}}{W_n}
\quad \mbox{ and } \quad
\bar{s}_{\rm tar}^{(\beta)} = 
\frac{\sum_{i,j \in V} w_{ij} s_j^{(\beta)}}{W_n}
\end{equation*}
are the weighted mean of the $\alpha$-type strength of the 
source nodes and the $\beta$-type strength of the target nodes, respectively,
with $W_n := \sum_{i,j \in V} w_{ij}$.
The value of $r_{\alpha,\beta}$ is between~$-1$ and~$1$.
An implementation of the weighted, directed assortativity is available in the
open-source \proglang{R} package \pkg{wdnet}~\citep{yuan2023generating}.

The directed assortativity coefficients provide a nuanced perspective on node
connection attributes, going beyond the general assortativity coefficients based
solely on total strength. In the context of IONs, assortativity assesses the
preference of one sector with certain sector-level feature channels products to
another sector with another sector-level feature. The features of the supplying
sector and the receiving sector do not have to be the same feature. We
consider here only two features in-strength and out-strength.
Positive assortativity coefficients signify assortative-mixing, suggesting
that nodes with higher strength tend to connect with similarly strong nodes. In
contrast, negative values indicate disassortative-mixing, where high-strength
nodes connect with weaker ones. For instance, a positive out-in assortativity
coefficient means sectors with significant out-strength tend to channel their
products to sectors with high in-strength. Conversely, a negative out-out
assortativity coefficient suggests sectors with large out-strength are more
inclined to direct their products to sectors with low out-strength. Other
assortativity types can be understood in a similar vein.

In empirical studies, assessing the uncertainty of sample assortativity
coefficients is as crucial as it is in simple correlation
analysis. Unfortunately, past analyses often neglected to incorporate
uncertainty measures~\citep[e.g.,][]{cerina2015world}, limiting the depth of
discussions about assortativity coefficients' temporal evolution. Some changes
over time that seem dramatic might actually be minor when considering confidence
intervals. To address this oversight, we calculated the standard errors of the
sample assortativity coefficients using a recently developed jackknife approach
tailored for networks~\citep{lin2020theoretical}. This method involves
repeatedly calculating the assortativity of networks obtained by omitting
one sector at a time, along with its associated edges, from the observed
network.

\begin{figure}[tbp]
\centering
\includegraphics[width=\textwidth]{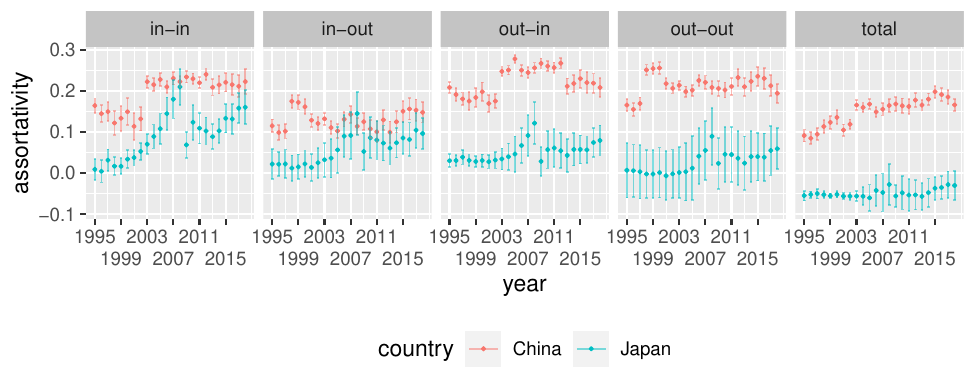}
\caption{Weighted and directed assortativity coefficients of the IONs 
  from China and Japan from 1995 to 2018. Error bars depict the range of one
  standard deviation above and below the sample assortativity coefficients,
  derived from a jackknife procedure.}
\label{fig:jackknife}
\end{figure}

Figure~\ref{fig:jackknife} showcases the assortativity coefficients for five
types, accompanied by their one-standard deviation error bars, for both China
and Japan from 1995 to 2018. A salient disparity in the assortativity
coefficient values between the two countries is evident. For China's IONs, all
five assortativity coefficients lie between 0.10 and 0.25. Given the minuscule
standard deviations resulting from the jackknife procedure, these coefficients
are distinctly positive. This positive assortative mixing indicates that sectors
in China with a higher out/in/total strength tend to direct their products more
towards sectors with a correspondingly high out/in/total strength, as opposed to
those with a lower strength.

In contrast, Japan presents a more varied landscape. While the in-in
assortativity coefficients are significantly positive post-2002, their
magnitudes are subdued compared to those of China, peaking in 2008 before
observing a marked decline-potentially an aftermath of the global financial
crisis. The in-out, out-in, and out-out coefficients, although positive, achieve
statistical significance only in 2008, followed by a substantial reduction. This
suggests a more nuanced, potentially non-obvious, assortative pattern in
Japan. The total assortativity coefficients for Japan even venture into the
negative spectrum, especially prominent up to 2002. Their modest absolute
values, around 0.05, hint at a slight inclination for high total-strength nodes
to associate with nodes of a lower total-strength. Further distinguishing the
two nations is the higher standard deviation of Japan's coefficients. Such a
variance indicates a more pronounced homogeneity in China's sectoral
connections: removing a single sector in China's network results in
comparatively minor shifts in its assortativity coefficients than in Japan.

The temporal progression of assortativity coefficients over the years brings
forth insightful revelations. Japan's IONs witnessed an uptrend in all directed
assortativity metrics until 2008, with in-in assortativity being particularly
pronounced. However, total assortativity remained relatively stable with a mild
negative inclination. Post-2008, a marked decline in directed assortativity
metrics is evident, a testament to the profound impact of the 2008 global
financial crisis on Japan. The crisis ostensibly disrupted the congruence of
sectoral supply chains and sectoral clusters within the
nation~\citep{grimes2009japan}. Given Japan's deep integration into global
economic operations, its susceptibility to international shocks became
manifestly evident.

Conversely, China's trajectory over the years offers a more intricate
tale. While the trends in the four directed assortativity metrics are
not obvious, there are exceptions that punctuate the narrative. Notably, 2003 saw a
spike in both in-in and out-in assortativities, likely a reverberation from
China's induction into the World Trade Organization two years prior. This
inclusion could have catalyzed an uptick in foreign demand, subsequently
fueling the expansion of domestic production scales and fortifying the supply
chain~\citep{andersen2014how}. Another conspicuous deviation
is observed in 1998 with a surge in in-out and out-out assortativities. This can
be traced back to China's nimble response to the 1997 Asian financial
crisis. Decisive interventions by the Chinese government, encompassing proactive
fiscal policies complemented by circumspect monetary strategies, invigorated
domestic demand. This likely culminated in more robust supply-centric sectoral
chain~\citep{wang1999theasian}. However, given that the
preponderant demand was steered by policy-induced infrastructural investments
during this epoch~\citep{li2000china}, this ascent proved ephemeral.

\section{Key sectors identification}
\label{sec:key}

Key sectors often underpin the growth and evolution of
economies~\citep{sonis1995linkages}. Recognizing their pivotal role, governments
may shape policies to bolster these sectors. However, this necessitates an
astute method for their identification, with many turning to the centrality
measures of IONs as a practical solution \citep[e.g.,][]{hewings1982empirical,
  depaolis2022identifying}. While a myriad of centrality definitions
exists~\citep[e.g.,][]{bonacich1987power, brin1998anatomy, newman2001scientific,
  barrat2004architecture}, it is vital to note that IONs are weighted and
directed networks. Moreover, they offer supplementary data beyond the
intermediate flow matrix, such as sector-level value added and final use
showcased in Table~\ref{tab:niot}. Therefore, a centrality measure that
accommodates both node weight and edge direction, supplemented by auxiliary IOT
variables, is desired.

We utilized an extended PR measure tailored for weighted, directed
networks, integrating auxiliary ranking information as detailed by
\citet{zhang2022pagerank}.
Specifically, the PR measure of node~$i$ is
\begin{equation*}
  P_i = \gamma \sum_{j \in V} \frac{w_{ji}}{s^{\text{(out)}}_j} P_j 
  + \frac{(1 - \gamma)\lambda_i}{\sum_{i \in V} \lambda_i},
  \qquad i = 1, \ldots, n,
\end{equation*}
where $\lambda_i$ is an auxiliary measure (which could be internal or external
to the network structure) of the relative importance of node~$i$, and
$\gamma \in [0, 1]$ tunes the relative importance of the auxiliary measure.
When all $\lambda_i$'s are equal, this measure reduces to a weighted
PR~\citep{ding2011applying}. When, additionally, $w_{ji}$ is replaced with
$a_{ji}$ and out-strength $s^{\text{(out)}}_j$ with out-degree, the measure
further reduces to the standard PR~\citep{page1998pagerank}. The influence of
the auxiliary variable $\lambda_i$ operates in a scale-free manner; the
contribution $\lambda_i$ to the PR lies in its proportional contribution to the
entire economy rather than its absolute value.

In the standard PR, $\gamma$ serves as a damping
factor that prevents the iteration for eigenvector in the algorithm from
getting stuck in sinking nodes (those without outgoing edges).
Typically, a value of $\gamma = 0.85$ is adopted, as recommended by
\citet{page1998pagerank}. In the context of the extended
PR~\citep{zhang2022pagerank}, a sector attains a high PR score if it possesses a
prominent auxiliary ranking, if it has substantial incoming edge weights, or if
its immediate upstream sectors exhibit high PR scores. Our investigations
indicated that setting $\gamma = 0.85$ strikes a desired balance between the ION
and the auxiliary metric. A lower $\gamma$, like 0.5, would overly emphasize
the auxiliary variable; see Section~1 in the Supplementary Material for a
comparison. The results reported next were obtained with $\gamma = 0.85$ as
implemented in the \proglang{R} package \pkg{wdnet}~\citep{yuan2023generating}.

\begin{figure}[tbp]
\centering
\includegraphics[width=\textwidth]{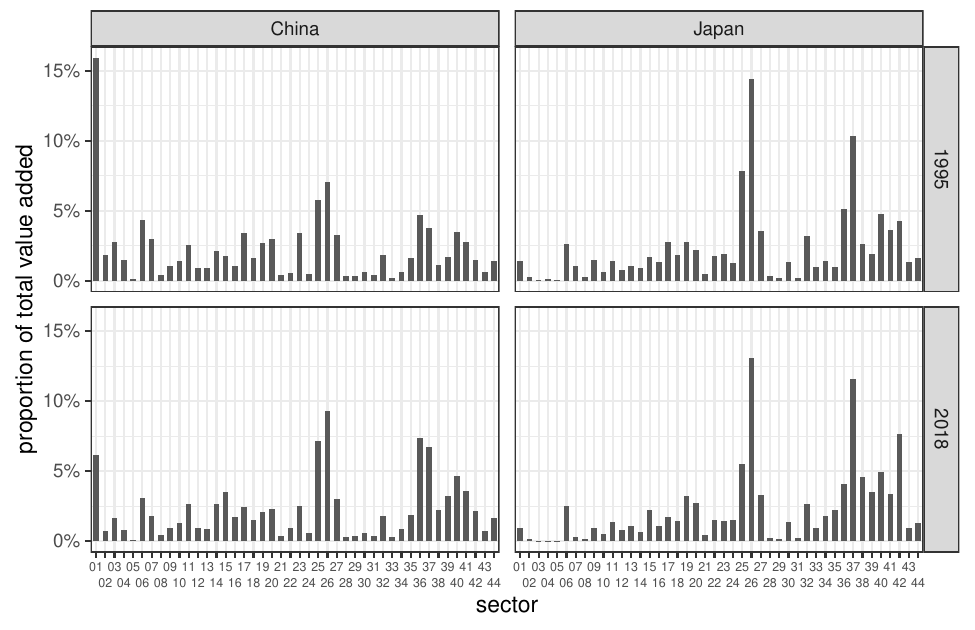}
\caption{Bar plots of the share of valued added of each sector in the GDP
in China and Japan in 1995 and 2018.}
\label{fig:tva}
\end{figure}

The selection of the auxiliary variable, represented by $\lambda_i$'s depends on
the research objectives. For instance, in labor economics, one might opt for
sectoral employment scale as the auxiliary data, whereas in innovation
economics, the sectoral research and development expenditure might be more apt,
as highlighted by \citet{xiao2022incorporating}. In our investigation, we used
the sectoral value added as the auxiliary metric~\citep{zhang2022pagerank}. This
choice was logical because the sectoral value added is readily accessible within
the IOT, yet remains external to the ION's formation. Additionally, this metric
adeptly reflects the economic prominence of each sector. To offer perspective,
Figure~\ref{fig:tva} displays the sectoral shares to the national GDP for 1995
and 2018 in China and in Japan. As expected, while Japan's service sectors have
higher value-added contributions, China's manufacturing sectors lead in this
regard. Nonetheless, the heterogeneity in sector-level value added is similar
for both nations. In 1995 and 2018, China's standard deviations were 0.0261 and
0.0214, and Japan's were 0.0275 and 0.0278. With a focus on economic growth,
integrating value added as an auxiliary insight when discerning key sector
rankings via PR scores offers a compelling analytical lens.

\begin{table}[tbp]
\centering
\caption{The sectors with top~5 extended PR scores of China and Japan from 1995
  to 2018 every three years with value added as auxiliary information.}
\label{tab:pr}
\begin{tabular*}{\textwidth}{@{\extracolsep{\fill}}ccccccccccc}
\toprule
Country & Rank & 1995 & 1998 & 2001 & 2004 & 2007 & 2010 & 2013 & 2016 & 
2018 \\ 
\midrule
China & 1 & 25 & 07 & 25 & 07 & 25 & 25 & 25 & 25 & 25 \\ 
& 2 & 06 & 06 & 07 & 25 & 07 & 17 & 20 & 06 & 17 \\ 
& 3 & 07 & 25 & 06 & 17 & 17 & 07 & 06 & 17 & 40 \\ 
& 4 & 26 & 01 & 01 & 06 & 15 & 19 & 26 & 20 & 26 \\ 
& 5 & 15 & 26 & 26 & 19 & 19 & 20 & 17 & 07 & 07 \\[1ex]
Japan & 1 & 26 & 26 & 26 & 26 & 26 & 26 & 26 & 26 & 26 \\ 
& 2 & 25 & 25 & 25 & 20 & 20 & 42 & 42 & 42 & 42 \\ 
& 3 & 42 & 42 & 42 & 42 & 42 & 20 & 20 & 20 & 20 \\ 
& 4 & 20 & 20 & 20 & 25 & 37 & 37 & 37 & 37 & 37 \\ 
& 5 & 37 & 37 & 37 & 37 & 25 & 25 & 25 & 25 & 25 \\ 
\bottomrule
\end{tabular*}
\end{table}

Table~\ref{tab:pr} presents the top~5 sectors in China and Japan ranked by their
PR scores from 1995 to 2018, with value added as auxiliary information. The
economic landscapes of these countries, as represented by these sectors, exhibit
stark contrasts, reflecting their individual development narratives and
strategic priorities. For China, a fluidity in the top sectors can be observed
over the years. Between 1995 and 2002, primary sectors such as ``agriculture,
hunting, forestry''~(01), ``food products, beverages and tobacco''~(06), and
``textiles, textile products, leather and footwear''~(07) dominated the list,
interspersed with ``construction''~(25) and ``wholesale and retail trade; repair of
motor vehicles''~(26). The prominence of sectors 07 and 25, which interchanged
leadership roles, underscores China's focus on both its agrarian roots and its
burgeoning infrastructure development. However, post-2007 marks a clear shift
towards industrialization, with the rise of manufacturing sectors like
``computer, electronic and optical equipment''~(17) and ``motor vehicles, trailers
and semi-trailers''~(20), while ``construction''~(25) retained its top position. This
transition can be attributed to China's strategic push towards rapid
industrialization, technological adoption, and urbanization
processes~\citep{zhang2014impact}.

In stark contrast, Japan exudes stability in its top sectors. The
unwavering dominance of ``wholesale and retail trade; repair of motor
vehicles''~(26) as the foremost sector signifies Japan's entrenched consumer and
trade dynamics. Furthermore, the consistent presence of ``construction''~(25),
``motor vehicles, trailers and semi-trailers''~(20), ``human health and social work
activities''~(42), and ``real estate activities''~(37) in the top ranks underlines
Japan's position as a mature, service-oriented, and industrialized economy,
aligning with its rich history as a global trade and technological
powerhouse~\citep{flath2022japanese}. We conjecture that such top-sector
stability might emerge at a particular phase of economic development when the
extended PR method leverages value-added as auxiliary information. China
might need decades to reach this evolutionary stage~\citep{nadvi1994industrial}.


Our analysis of key sectors, when juxtaposed with the findings of
\citet{li2017examining}, presents both alignments and discrepancies. Both
studies recognize the dominance of manufacturing sectors in China, indicative of
its ongoing industrialization, and the crucial role of Japan's service sectors,
especially commerce and commercial service. This is reflected in the elevated PR
scores in Japan's sectoral network. Using degree centrality,
\citet{li2017examining} suggested diminishing prominence of the construction
sector in China during 2000--2011, possibly due to the bursting of an economic
bubble. Our study from the extended PR scoring points to its consistent
dominance over the years. Using betweenness centrality, \citet{li2017examining}
reported high ranks of the chemical sector and the nonmetallic mineral products
sector, followed by the metal smelting sector in Japan; and higher ranks of the
printing sector and the electronic manufacturing in 2002 and 2007 in
China. These different rankings are expected given the that the betweenness
centrality measures the ability of sectors in connecting distant
sectors~\citep{xiao2022incorporating}.

To enhance the depth of our analysis, we also used alternative centrality
metrics. Still with the extended PR score, we used export value as the
node-level auxiliary variable, recognizing its accessibility within IOTs and its
externality to the IONs. Emphasizing export value highlights sectors critical
for export-driven economic strategies, aligning with policy directives for
countries emphasizing export-led growth. This examination is detailed in
Appendix~\ref{sec:expo}. In addition, we applied a weighted version of the hubs
and authorities centrality, as per \citet{agosti2005theoretical} and the
foundational work of \citet{kleinberg1999authoritative} and
\citet{kleinberg1999hubs}. This metric, while more objective, omits auxiliary
data considerations but did spotlight additional vital sectors, further
discussed in Appendix~\ref{sec:hubau}.

\section{Community detection}
\label{sec:community}

Sectoral clusters are important for promoting national competitiveness,
structural change, and economic development
\citep[e.g.,][]{learmonth2003multi, titze2011identification}. These clusters
consist of sectors that are more closely connected to one another than with
other sectors, and community detection in a network can group nodes into
clusters or communities based on their bond strength. Many algorithms for
community detection are available
\citep[e.g.,][]{ng2001spectral, girvan2002community, boykov2004experimental}.
In this study, we used the modularity maximization algorithm developed by
\citet{newman2006modularity} to detect communities in a weighted network
The modularity $Q$ of a weighted network is
\begin{equation*}
Q = \frac{1}{W_n} \sum_{ij}
\left( w_{ij} - \frac{s_i s_j}{W_n} \right) \mathbb{I}(h_i = h_j),
\end{equation*}
where $h_i$ denotes the community to which node $i$ is assigned, 
and $\mathbb{I}(\cdot)$ is the indicator function taking value 1 
if nodes~$i$ and~$j$ are in the same community and 0 otherwise.
For node~$i$ and~$j$, the term $\left( w_{ij} - s_i s_j/W_n \right)$ represents
the difference between the actual edge weight and what would be expected if all
the edges are randomly placed among the nodes. From the definition, higher
$Q$~values reflects more pronounced community structure of the network 
under the corresponding clustering strategy ${\bm h} := (h_i)_{i = 1}^{n}$. 
Therefore, community detection aims to maximize $Q$ with respect to all
possible clustering strategies. We used the greedy algorithm 
proposed by~\citet{clauset2004finding} to solve the optimization problem,
which is provided in the \proglang{R} package
\pkg{igraph}~\citep{csardi2006igraph}.

\begin{figure}[tbp]
  \centering
  \includegraphics[width=\textwidth]{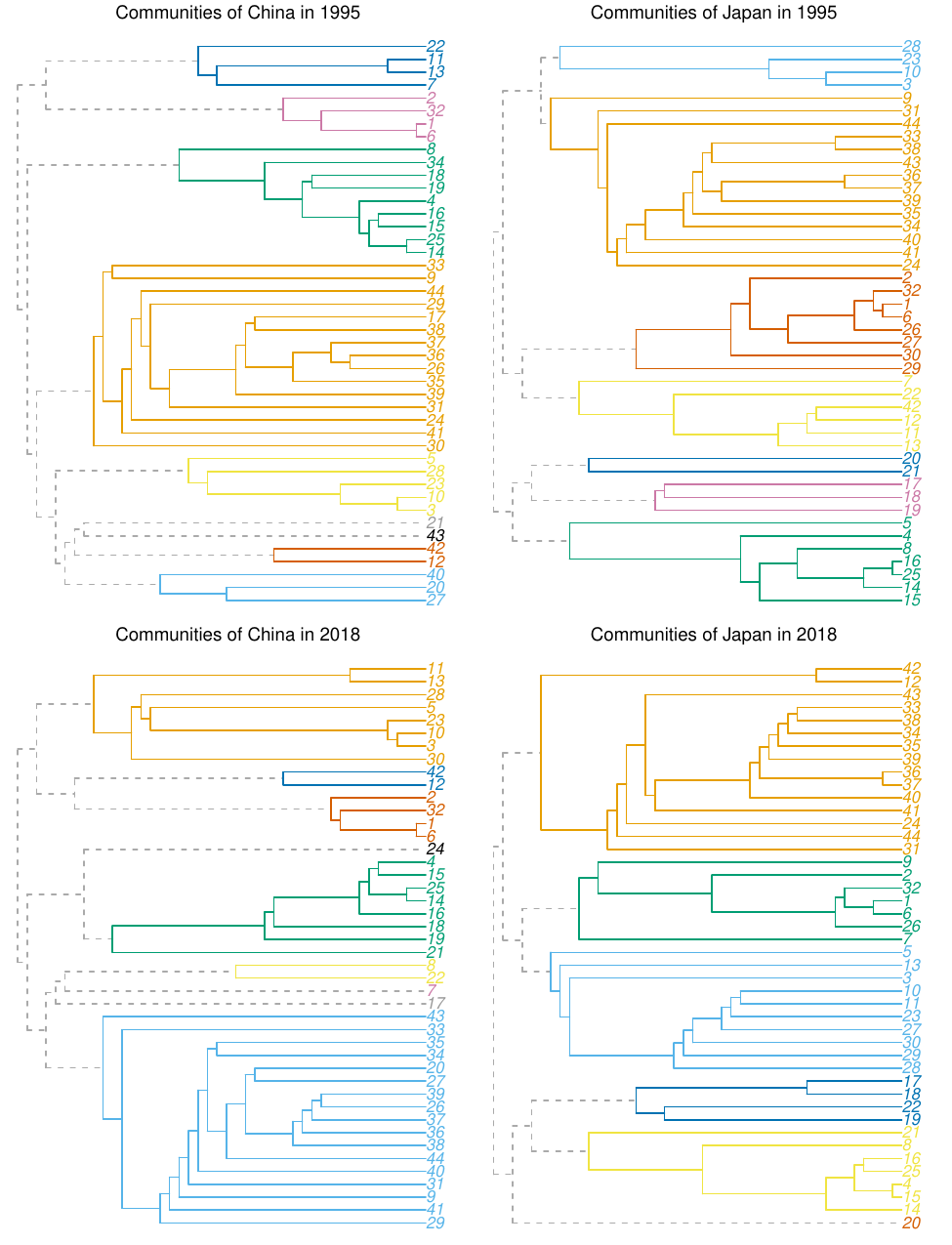}
  \caption{Community detection of China and Japan in 1995 and 2018.}
  \label{fig:cluster}
\end{figure}

Figure~\ref{fig:cluster} displays the community structures of China and Japan in
1995 and 2018; results from other years are available in Section 2 of the
Supplementary Material.
Some community structures clearly reflect different parts of sectoral
chains including upstream, midstream, and downstream sectors.
For instance, one community of China in 1995 includes 
``agriculture, hunting, forestry''~(01), ``fishing and aquaculture''~(02), 
``food products, beverages and tobacco''~(06), and 
``accommodation and food service activities''~(32). 
This community covers the complete life cycle of agricultural products, 
from planting to processing to consumption. 
To be specific, as the upstream sectors, 
``agriculture, hunting, forestry''~(01) and ``fishing and aquaculture''~(02) 
provide raw materials for agricultural products. By acquiring raw materials for 
reprocessing, the midstream sector ``food products, beverages and tobacco''~(06)
carries out the production and then acts as the bridge between upstream and 
downstream, moving the products to the downstream sector 
``accommodation and food service activities''~(32), 
which is the consumer of food products. 
Another example is the community of Japan in 1995, which presents a clear 
picture of the production and consumption process of the medical industry, 
including ``chemical and chemical products''~(11), 
``pharmaceuticals, medicinal chemical and botanical products''~(12), 
``rubber and plastics products''~(13), and 
``human health and social work activities''~(42).
Additionally, both countries possess construction-centered communities, which 
also include upstream sectors for the ``construction''~(25), such as 
``other non-metallic mineral products''~(14), ``basic metals''~(15), and 
``fabricated metal products''~(16).

The community structures of China and Japan show differences in the formation of
the communities. In Japan, the top sector
``wholesale and retail trade; repair of motor vehicles''~(26) usually appears
with ``agriculture, hunting, forestry''~(01), 
``food products, beverages and tobacco''~(06), and 
``accommodation and food service activities''~(32) in the same community.
In China, however,
``wholesale and retail trade; repair of motor vehicles''~(26) is not
in the same community as food production and services, but in the 
service-centered community. In Japan, sectors ``electrical equipment''~(18) and
``machinery and equipment, not elsewhere classified''~(19) are in the same
community with sectors ``computer, electronic and optical equipment''~(17) or 
``other transport equipment''~(21), which demonstrates the clustering
characteristic of the Japan's transport equipment manufacturing industry.
In China, however, the two sectors mostly belong to the 
construction-centered community.
Another interesting example is about the biopharmaceutical industry. China's 
``pharmaceuticals, medicinal chemical and botanical products''~(12) and 
``human health and social work activities''~(42) are closely linked to belong 
to the same community, but do not form a stable community structure with 
``chemical and chemical products''~(11) and 
``rubber and plastics products''~(13) as they do in Japan, implying 
there was a quite gap between the two countries in biopharmaceutical industry. 
In the Japanese service sector community,
``financial and insurance activities''~(36) and ``real estate activities''~(37) 
are closely linked, illustrating the connection between Japan's real estate and 
financial industries. While there is no apparent clustering between the two 
sectors in China's service sector community.

Over the 24-year period, the community structure within each country has
undergone changes, but certain sectors have remained consistent within the same
community. In China, the sectors of ``agriculture, hunting, forestry''~(01),
``fishing and aquaculture''~(02), ``food products, beverages and tobacco''~(06), and
``accommodation and food service activities''~(32) have remained in the same
community between 1995 and 2018. Similarly, the sectors of ``mining and
quarrying, energy producing products''~(03), ``coke and refined petroleum
products''~(10), and ``electricity, gas, steam and air conditioning supply''~(23)
have also been relatively stable in their community affiliations. Likewise,
there are certain sectors in Japan that have exhibited consistent community
structures. For example, the community composed of ``agriculture, hunting,
forestry''~(01), ``food products, beverages and tobacco''~(06), ``wholesale and
retail trade; repair of motor vehicles''~(26), and ``accommodation and food
service activities''~(32) has remained relatively stable over time.
The 24-year period of the study may not have been
long enough to observe drastic changes in community structures. However, the
consistent presence of certain sectors within the same community across the
study period suggests that these sectors have remained interconnected and
interdependent over time.

\begin{figure}[tbp]
\centering
\includegraphics[width=\textwidth]{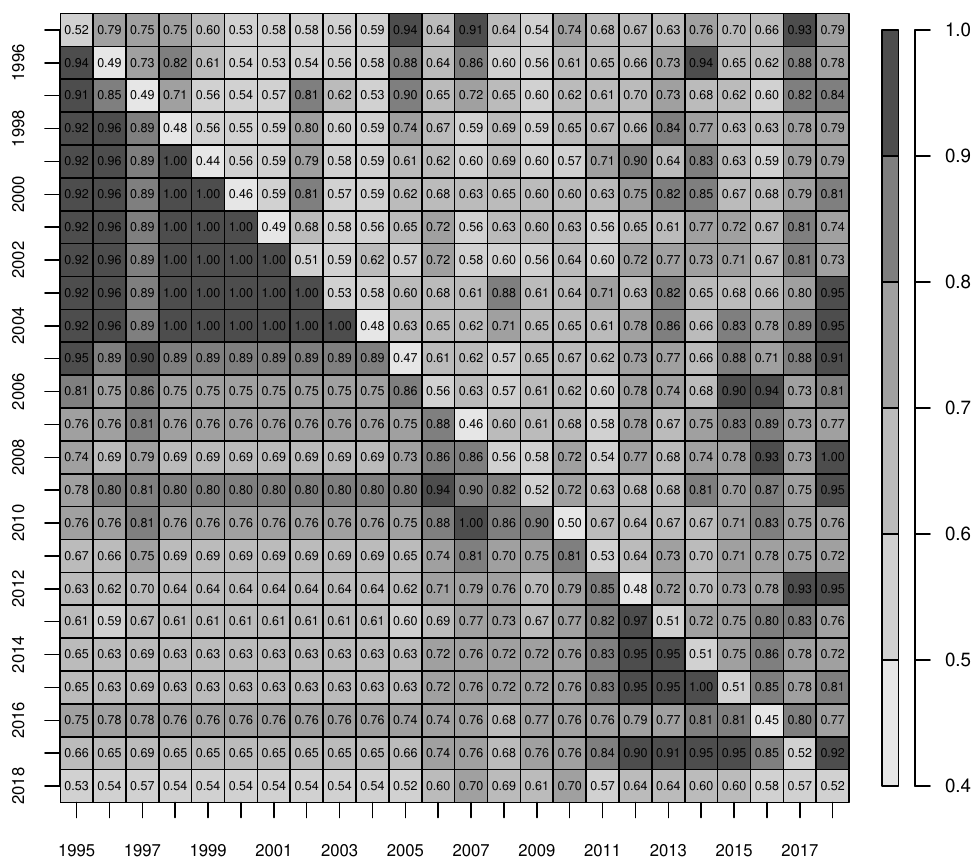}
\caption{The AMI in the community results of China and Japan from 1995 to 2018.
  The upper triangle entries indicate the pairwise similarities of community
  structures within China over time. The lower triangle entries reflect the
  pairwise similarities of community structures within Japan over time. The
  diagonal entries are the similarities between China and Japan over time.}
\label{fig:AMI}
\end{figure}


The evolvement of the community structures over time can be quantitatively
demonstrated by similarity measures between any pair of community structures.
In particular, we used adjusted mutual information
(AMI)~\citep{vinh2009information,vinh2010information} as our similarity measure.
The AMI takes value~1 when two identical community structures and~0  when the
mutual information between two community structures equals the value expected
due to chance alone. A higher AMI value indicates a higher agreement between
the two community structures. Figure~\ref{fig:AMI} visualizes the year-by-year
AMI matrix whose elements are pairwise AMI values. The diagonal entries capture
annual comparisons between the community structures of the two countries over
the years. The fluctuating AMI values around $0.5$ suggests a moderate level of
similarity between the community structures in China and in Japan.
The upper and lower triangle entries are year-year similarities
in China's and Japan's community structures, respectively.
In both countries, community structures exhibit similarities over~$0.5$.
Japan's structures follow a consistent pattern, with high similarities
before 2005 that gradually decreases over time. This decline might be
attributed to minor shifts likely caused by advanced industrial development.
China's community structures over time, which are less similar than
Japan's. In addition, there appears to be no clear pattern. Even between
adjacent years, the AMI values are rarely above~$0.8$. This possibly stemmed
from its rapid economic growth and significant changes in sectoral structure. A
full exploration into the forces steering these dynamics would be a compelling
topic for subsequent research endeavors.

\section{Concluding remarks}
\label{sec:dis}

In this comparative study of China and Japan's sectoral structures, we applied
several network analysis techniques, including some that were recently
developed. We began with an examination of node strength distributions and
illustrated sectoral connections. For a more nuanced understanding of the
homophily of the supplying sectors and receiving sectors, we employed weighted,
directed assortativity coefficients~\citep{yuan2021assortativity}, considering
uncertainties through the jackknife method~\citep{lin2020theoretical}. Without
uncertainty quantifications, some conclusions in existing works could be
misleading \citep{cerina2015world}. To
identify key sectors, we used the extended PR method that integrates
research-specific auxiliary metrics, like total value added or export
values~\citep{zhang2022pagerank}. We also detected sectoral community
structures within each economy and assessed their similarities, both between
countries and over time, using the AMI metric~\citep{vinh2009information,
  vinh2010information}. With this combination of methods, we provided detailed
insights into the evolving sectoral dynamics of two pivotal Asian economies.


Utilizing network centrality measures to pinpoint key sectors in IONs is a
logical approach, and there are hundreds of available centrality
measures~\citep[e.g.,][]{perra2008spectral, landherr2010critial, das2018study}.
The choice of a particular measure need to align with the unique objectives set
by the researchers. While we leveraged the extended PR
method~\citep{zhang2022pagerank} and hubs and
authorities~\citep{deguchi2014hubs}, there exist other commonly used centrality
metrics such as degree, betweenness, closeness, and random-walk based measures
that have been applied to ION analyses~\citep{blochl2011vertex, li2017examining,
	tsekeris2017network, depaolis2022identifying}. The key is not to deem
one method superior to another but to select based on the specific research
context. Our preference for the extended PR method was due to its
adaptability in accounting for external factors, allowing for a tailored
approach depending on the research objective. For instance, researchers
examining the interplay between industrial growth and employment might opt to
integrate employment data as auxiliary
input~\citep{xiao2022incorporating}. The diversity of centrality
measures offer rich opportunities and perspectives to identify key sectors in
IONs.


As an important tool for economic system research, IOTs are available from
several sources. While national IOTs composed by individual countries contain
detailed view of sectoral structures tailored to their specifications,
inconsistencies in sector classifications hinder country comparisons. To
facilitate comparison, some international organizations composed IOTs. Some of
these contains individual national IOTs such as STAN,
while others cover amalgamate data from multiple countries into expansive
multi-region IOTs, such as WIOD and
ADB-MRIO. With each country as a block, the
diagonal blocks of a multi-region IOT give individual national IOTs. Our study
did not use the WIOD due to its timeline ending in 2014, nor the ADB-MRIO data
base because incorporates two non-applicable sectors for China. Nevertheless,
future studies might benefit from analyzing the WIOD and data from other
sources, such as the Eora Global Supply Chain
Database~\citep{lenzen2013building} and the Global Trade Analysis
Project~\citep{aguiar2022global}, and juxtaposing the findings with ours for
overlapping years. Although the 45-sector granularity of
the STAN database serves our purposes, a finer sectoral resolution aligning the
sectors of both China's and Japan's IOTs, which respectively have 142 and 187
sectors, would deliver deeper and more intricate perspectives.





Multi-region IOTs encompass not only the interconnections among sectors within
individual countries but also the economic transactions between the sectors across
nations~\citep{dietzenbacher2013construction, lenzen2013building,
  aguiar2022global}. Such databases are essential 
for conducting research in the field of international economics. By harnessing 
international IOTs in conjunction with network analysis methods, 
we can delve into various facets of economic relationships among different 
countries. This includes examining international 
trade~\citep{zhu2014rise,cerina2015world}, intermediate and final
goods trade~\citep{kleinert2003growing},
global supply chains and value chains~\citep{angelidis2020competitive},
energy flows~\citep{chen2018global}, and carbon emission~\citep{minx2009input},
among others. These insights hold significant implications for the formulation
of trade, energy, and environmental policies. They provide a robust framework
for assessing the stability of supply chains, gauging economic vulnerabilities,
and evaluating the resilience of nations. Particularly concerning China-Japan
economic relations, there remains a myriad of research questions related to their
economic interplay that can be adeptly tackled using multi-region IOTs in tandem
with advanced network analysis methodologies.

Impacts of global shocks to sectoral structures present a rich vein of
research. Notably, the 2008 global financial crisis has been analyzed for its
ripple effects on global economies and network
structures~\citep{haldane2013rethinking, battiston2012debtrank}. While
our study touched upon the repercussions of the 2008 crisis in China and Japan,
a deeper exploration through a network lens is warranted. The
Covid-19 pandemic, an unparalleled recent shock, has also been a focal point of
many economic studies~\citep{baldwin2020economics, brodeur2021literature,
  aktar2021global, costa2022sectoral, tanaka2022economic, salisu2022note,
  ozili2023spillover}. The
updated Eora Global Supply Chain Database~\citep{lenzen2013building} provides an
excellent opportunity to delve into the pandemic's impact on economies like
China and Japan~\citep{han2022impact, kitamura2020evaluation}. Such
explorations could illuminate the resilience and
adaptability of these nations when faced with monumental challenges.

\section*{Acknowledgments}
TW and SX were supported by the National Social Science Fund of China (21BTJ013)
and the National Bureau of Statistics of China (2022LY095).
TW was supported by Shanxi Scholarship Council of China  (2023-118).
JY's research was partially supported by NSF grant DMS2210735.

\appendix

\section{Sector codes}
\label{sec:seccode}

Table~\ref{tab:stan} summarizes the codes and the definitions of the 44 sectors 
in the 2021 edition of STAN database~\citep{OECD2021stand}. The 45th sector,
``activities of households as employers; undifferentiated goods- and
services-producing activities of households for own use'', is missing for both
China and Japan.

\begin{table}[tbp]
\centering
\caption{Description of the sectors in the STAN.}
\label{tab:stan}
\begin{tabular}{cp{.4\textwidth}cp{.4\textwidth}}
\toprule
Code & \multicolumn{1}{c}{Sector} & Code & \multicolumn{1}{c}{Sector} \\
\midrule
01 & Agriculture, hunting, forestry & 23 & Electricity, gas, steam and 
air conditioning supply \\ 
02 & Fishing and aquaculture & 24 & Water supply; sewerage, waste 
management and remediation activities \\ 
03 & Mining and quarrying, energy producing products & 25 & 
Construction \\ 
04 & Mining and quarrying, non-energy producing products & 26 & 
Wholesale and retail trade; repair of motor vehicles \\ 
05 & Mining support service activities & 27 & Land transport and 
transport via pipelines \\ 
06 & Food products, beverages and tobacco & 28 & Water transport \\ 
07 & Textiles, textile products, leather and footwear & 29 & Air 
transport \\ 
08 & Wood and products of wood and cork & 30 & Warehousing and support 
activities for transportation \\ 
09 & Paper products and printing & 31 & Postal and courier activities 
\\ 
10 & Coke and refined petroleum products & 32 & Accommodation and food 
service activities \\ 
11 & Chemical and chemical products & 33 & Publishing, audiovisual and 
broadcasting activities \\ 
12 & Pharmaceuticals, medicinal chemical and botanical products & 34 & 
Telecommunications \\ 
13 & Rubber and plastics products & 35 & IT and other information 
services \\ 
14 & Other non-metallic mineral products & 36 & Financial and insurance 
activities \\ 
15 & Basic metals & 37 & Real estate activities \\ 
16 & Fabricated metal products & 38 & Professional, scientific and 
technical activities \\ 
17 & Computer, electronic and optical equipment & 39 & Administrative 
and support services \\ 
18 & Electrical equipment & 40 & Public administration and defence; 
compulsory social security \\ 
19 & Machinery and equipment, not elsewhere classified & 41 & Education 
\\ 
20 & Motor vehicles, trailers and semi-trailers & 42 & Human health and 
social work activities \\ 
21 & Other transport equipment & 43 & Arts, entertainment and 
recreation \\ 
22 & Manufacturing nec; repair and installation of machinery and 
equipment & 44 & Other service activities \\ 
\bottomrule
\end{tabular}
\end{table}

\section{Extended PR incorporating exports}
\label{sec:expo}

\begin{figure}[tbp]
\centering
\includegraphics[width=\textwidth]{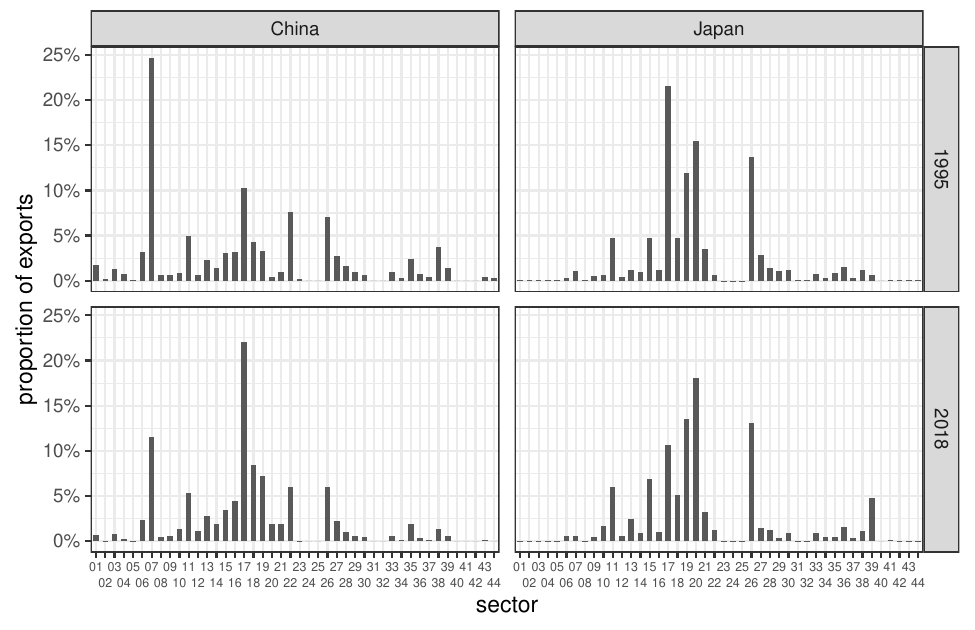}
\caption{Bar plots of the share of export value of each sector in the total
  value of exports in China and Japan for 1995 and 2018.}
\label{fig:expo}
\end{figure}

The extended PR allows flexible incorporation of auxiliary information in
centrality ranking depending on the research objectives
\citep{zhang2022pagerank}. Here we use export value as the sector-specific
auxiliary information in the extended PR. This variable is a component of the
final use of an IOT, as presented in Table~\ref{tab:niot}. Similar to value
added, it is also external to the construction of the corresponding ION. Export
value highlights sectors critical for export-driven economic strategies,
aligning with policy directives for countries emphasizing export-led growth.
Figure~\ref{fig:expo} presents the bar plots of the sectoral shares in the
national exports for 1995 and 2018 in China and in Japan.

\begin{table}[tbp]
\centering
\caption{The sectors with top~5 extended PR scores of China and Japan from 1995
  to 2018 every three years with export value as auxiliary information.}
\label{tab:prexpo}
\begin{tabular*}{\textwidth}{@{\extracolsep{\fill}}ccccccccccc}
\toprule
Country & Rank & 1995 & 1998 & 2001 & 2004 & 2007 & 2010 & 2013 & 2016 & 
2018 \\ 
\midrule
China & 1 & 07 & 07 & 07 & 17 & 17 & 17 & 25 & 25 & 25 \\
& 2 & 25 & 17 & 17 & 07 & 07 & 07 & 17 & 17 & 17 \\
& 3 & 26 & 25 & 25 & 25 & 25 & 25 & 07 & 07 & 07 \\
& 4 & 19 & 06 & 06 & 19 & 19 & 19 & 19 & 19 & 19 \\
& 5 & 17 & 26 & 26 & 15 & 15 & 20 & 20 & 20 & 20 \\ [1ex]
Japan & 1 & 20 & 20 & 20 & 20 & 20 & 20 & 20 & 20 & 20 \\
& 2 & 26 & 26 & 26 & 26 & 26 & 26 & 26 & 26 & 26 \\
& 3 & 17 & 17 & 17 & 17 & 17 & 17 & 19 & 19 & 19 \\
& 4 & 25 & 25 & 19 & 19 & 19 & 42 & 42 & 42 & 42 \\
& 5 & 19 & 19 & 25 & 25 & 15 & 19 & 25 & 25 & 25 \\
\bottomrule
\end{tabular*}
\end{table}

Table~\ref{tab:prexpo} summarizes top five sectors in China and Japan in their
extended PR scores, with export value as auxiliary information, from 1995 to
2018. In China, a clear prominence of the ``textiles,
textile products, leather and footwear''~(07) sector is observed at the onset,
which later gives way to sectors such as ``computer, electronic and optical
equipment''~(17) and ``construction''~(25) in subsequent years. The ``wholesale
and retail trade; repair of motor vehicles''~(26) and ``machinery and equipment,
not elsewhere classified''~(19) sectors persistently remain in the top five over the
years. For Japan, the tableau is marked by a steadfast dominance of the ``motor
vehicles, trailers and semi-trailers''~(20) sector, with ``wholesale and retail
trade; repair of motor vehicles''~(26) consistently following suit. Other
sectors like ``computer, electronic and optical equipment''~(17) and
``construction''~(25) maintain their relevance in the top five throughout the
years.

Contrasting these findings with the results from the value added auxiliary
information, we note divergent trends. In the value added scenario, China's
``construction''~(25) sector dominated, reflecting the nation's infrastructural
growth. However, when export value takes precedence, traditional manufacturing
sectors gain traction, underscoring the nation's export-oriented economic
growth. Japan, on the other hand, showcased more stability across both
scenarios, resulting from its mature, service-oriented economy. The
``human health and social work activities''~(42) sector, which consistently
appeared in Table~\ref{tab:pr}, shows up only post-2007. The discernable
shifts in sector prominence when transitioning from value-added to export value
highlight capability of the extended PR score in reflecting the multifaceted
nature of economic development.

\section{Weighted hubs and authorities}
\label{sec:hubau}

Hubs and authorities are two centrality measures involved in hyperlink-induced
topic search (HITS) algorithm to rank webpages
\citep{kleinberg1999authoritative,kleinberg1999hubs}. The key idea is that web
pages serve two primary roles: they provide content (making them an authority)
and they link to other pages (making them a hub). A good authority is a page
that is linked to by many good hubs. A good hub is a page that links out to many
good authorities. The two concepts are interdependent. A hub's value is
determined by the quality of the authorities it links to. An authority's value
is determined by the quality of the hubs that link to it. The HITS algorithm
involves a recursive process: it begins with an arbitrary assignment of values
to nodes as hubs and authorities. Then, it updates the hub and authority scores
of each node based on the initial (or previous) scores of its neighboring
nodes. This iterative process continues until the scores converge. In theory,
the hub score and the authority score eventually converge to the principal
eigenvectors of $AA^\top$ and $A^\top A$, respectively, where $A$ is the
adjacency
matrix of the network. To stabilize the numerical range and facilitate the
comparison of the relative importance, the results are normalized to make the
two score vector sums equal to one.

The HITS algorithm has been extended to allow edge weight by replacing the
adjacency matrix $A$ with the weighted adjacency matrix $W$ in the
calculation~\citep{agosti2005theoretical}. The weighted HITS algorithm finds
applications in a variety of fields, such as trade
network~\citep{deguchi2014hubs} and road network~\citep{sun2018role}.
We used the implementation in \proglang{R} package
\pkg{igraph}~\citep{csardi2006igraph} in our analysis.

\begin{table}[tbp]
\centering
\caption{The sectors with top~5 weighted hub and authority centrality scores of
China and Japan from 1995 to 2018 every three years.}
\label{tab:hits}
\begin{tabular*}{\textwidth}{@{\extracolsep{\fill}}cccccccccccc}
\toprule
Centrality & Country & Rank & 1995 & 1998 & 2001 & 2004 & 2007 & 2010 & 2013 & 
2016 & 2018 \\
\midrule
Hub & China & 1 & 15 & 15 & 15 & 15 & 15 & 15 & 15 & 15 & 15 \\ 
& & 2 & 07 & 07 & 14 & 14 & 14 & 14 & 14 & 01 & 14 \\ 
& & 3 & 01 & 11 & 01 & 01 & 17 & 04 & 11 & 14 & 01 \\ 
& & 4 & 14 & 01 & 26 & 11 & 04 & 11 & 01 & 06 & 26 \\ 
& & 5 & 26 & 14 & 11 & 26 & 11 & 19 & 26 & 26 & 17 \\ [1ex]
& Japan & 1 & 26 & 26 & 26 & 20 & 15 & 15 & 15 & 20 & 20 \\ 
& & 2 & 20 & 20 & 20 & 26 & 20 & 26 & 26 & 26 & 15 \\ 
& & 3 & 15 & 15 & 15 & 15 & 26 & 20 & 20 & 15 & 26 \\ 
& & 4 & 16 & 16 & 16 & 19 & 19 & 19 & 03 & 38 & 19 \\ 
& & 5 & 14 & 19 & 19 & 16 & 38 & 38 & 38 & 39 & 38 \\ [2ex]
Authority & China & 1 & 07 & 07 & 25 & 15 & 15 & 15 & 25 & 25 & 25 \\ 
& & 2 & 25 & 25 & 15 & 25 & 25 & 25 & 15 & 06 & 15 \\ 
& & 3 & 15 & 15 & 19 & 19 & 19 & 19 & 06 & 15 & 17 \\ 
& & 4 & 06 & 06 & 06 & 17 & 17 & 18 & 11 & 01 & 06 \\ 
& & 5 & 19 & 11 & 07 & 18 & 18 & 17 & 19 & 07 & 07 \\ [1ex]
& Japan & 1 & 25 & 25 & 20 & 20 & 15 & 15 & 15 & 20 & 20 \\ 
& & 2 & 20 & 20 & 25 & 25 & 20 & 20 & 20 & 15 & 15 \\ 
& & 3 & 15 & 26 & 26 & 26 & 19 & 26 & 26 & 26 & 26 \\ 
& & 4 & 26 & 06 & 06 & 15 & 25 & 25 & 25 & 25 & 25 \\ 
& & 5 & 06 & 19 & 19 & 19 & 26 & 19 & 19 & 19 & 19 \\ 
\bottomrule
\end{tabular*}
\end{table}

In IONs, sectors that act as major hubs distribute products or services to
various authority sectors, while high-authority sectors receive from many
notable hubs. The HITS method, unlike the extended PR
approach~\citep{zhang2022pagerank}, does not incorporate auxiliary information,
making it straightforward but potentially omitting valuable data. Hub and
authority scores showcase distinct facets of sectoral significance within the
IONs. Table~\ref{tab:hits} lists the top five sectors by weighted hub and
authority scores from 1995 to 2018 for China and Japan. As expected, the results
are different from those obtained using the PR method with value added as
auxiliary information.

The distinct trajectories of China and Japan are evident in their leading hub
sectors. China's consistent top hub sectors like ``basic metals''~(15), ``other
non-metallic mineral products''~(14), and ``agriculture, hunting,
forestry''~(01) across the years reveal its industrial foundation, rooted in
heavy industries and primary sectors. This mirrors China's expansive
manufacturing base and the momentum of its rapid urbanization. Conversely,
Japan's hub sectors gravitate towards ``wholesale and retail trade; repair of
motor vehicles''~(26), ``basis metals''~(15), and ``motor vehicles, trailers and
semi-trailers''~(20), underscoring its pivot towards export-driven industries,
notably in technology and automobiles. This orientation aligns with Japan's
established trade relationships and its globally esteemed automotive and tech
brands.

The leading authorities also demonstrate varying sectoral strengths and
priorities. China's top authority sectors, including ``textiles, textile
products, leather and footware''~(07), ``basic metals''~(15), and
``construction''~(25) in subsequent years, highlight its robust infrastructure
expansion, fueled by industrial policies that champion infrastructure as an
economic growth catalyst. In contrast, Japan's authority sectors such as
``construction''~(25), ``motor vehicles, trailers and semi-trailers''~(20), and
``basic metals''~(15) reflect its stature as a developed, industrialized nation,
renowned as a global automobile leader, and underscore its consumer-centric
market dynamics. These sectors' enduring relevance attests to Japan's
sophisticated, service-led economy.

\bibliographystyle{chicago}
\bibliography{cjiot}

\end{document}


\maketitle

\section{Extended PR with different tuning parameters}

\begin{table}[tbp]
\centering
\caption{The sectors with top~5 extended PR scores,
based on the sensitivity analysis of the tuning parameter at the levels
$\gamma = \{0.5, 0.85\}$ for China and Japan from 1995 to 2018
every three years with value added as auxiliary information.}
\label{tab:prsens}
\resizebox{\textwidth}{!}{
\begin{tabular}{cccccccccccccccccccc}
\toprule
\multirow{2}{*}{Country} & \multirow{2}{*}{Rank}
& \multicolumn{9}{c}{$\gamma = 0.5$} & \multicolumn{9}{c}{$\gamma = 0.85$} \\
\cmidrule(lr){3-11} \cmidrule(lr){12-20} 
& & 1995 & 1998 & 2001 & 2004 & 2007 & 2010 & 2013 & 2016 & 2018 & 
1995 & 1998 & 2001 & 2004 & 2007 & 2010 & 2013 & 2016 & 2018 \\ 
\midrule
China & 1 & 01 & 01 & 01 & 01 & 25 & 25 & 25 & 25 & 25 & 25 & 07 & 25 & 07 & 25 
& 25 & 25 & 25 & 25 \\ 
& 2 & 06 & 06 & 25 & 25 & 01 & 01 & 26 & 26 & 26 & 06 & 06 & 07 & 25 & 07 & 17 
& 20 & 06 & 17 \\ 
& 3 & 26 & 25 & 06 & 06 & 26 & 26 & 01 & 01 & 01 & 07 & 25 & 06 & 17 & 17 & 07 
& 06 & 17 & 40 \\ 
& 4 & 25 & 26 & 26 & 26 & 06 & 06 & 06 & 06 & 40 & 26 & 01 & 01 & 06 & 15 & 19 
& 26 & 20 & 26 \\ 
& 5 & 07 & 07 & 07 & 07 & 07 & 17 & 36 & 36 & 36 & 15 & 26 & 26 & 19 & 19 & 20 
& 17 & 07 & 07 \\ [1ex]
Japan & 1 & 26 & 26 & 26 & 26 & 26 & 26 & 26 & 26 & 26 & 26 & 26 & 26 & 26 & 26 
& 26 & 26 & 26 & 26 \\ 
& 2 & 25 & 25 & 37 & 37 & 37 & 37 & 37 & 37 & 37 & 25 & 25 & 25 & 20 & 20 & 42 
& 42 & 42 & 42 \\ 
& 3 & 37 & 37 & 25 & 42 & 42 & 42 & 42 & 42 & 42 & 42 & 42 & 42 & 42 & 42 & 20 
& 20 & 20 & 20 \\ 
& 4 & 42 & 42 & 42 & 25 & 25 & 25 & 25 & 25 & 25 & 20 & 20 & 20 & 25 & 37 & 37 
& 37 & 37 & 37 \\ 
& 5 & 36 & 36 & 36 & 36 & 20 & 20 & 20 & 20 & 20 & 37 & 37 & 37 & 37 & 25 & 25 
& 25 & 25 & 25 \\  
\bottomrule
\end{tabular}}
\end{table}


The choice of the parameter $\gamma$ in the context of extended PR
rankings~\citep{zhang2022pagerank} holds substantial importance, as it
determines the balance between the weight placed on the ION structure and the
auxiliary information, in this case, value-added, depending on the researchers'
objectives and preferences. Although $\gamma = 0.5$ suggests an even weighting
between the network structure and the value-added, it over-dilutes the influence
of the network's connectivity when deriving significance rankings for sectors
based on the IONs. Table~\ref{tab:prsens} presents a side-by-side comparison of
the top~5 sector for China and Japan from the extended PR algorithm using
value added as auxiliary information, with tuning parameters $\gamma = 0.5$ and
$\gamma = 0.85$ from 1995 to 2018, assessed every three years.

In China, at $\gamma = 0.5$, the sectors like ``agriculture, hunting,
forestry''~(01) and ``textiles, textile products, leather and footwear''~(07)
are predominant across several earlier years. But as $\gamma$ shifts to 0.85,
which puts more emphasis on the network structure, there is a marked transition
in the rankings. Specifically, the prominence of the ``construction''~(25)
sector becomes apparent and early, while the significance of the ``agriculture,
hunting, forestry''~(01) sector diminishes. The rankings from $\gamma = 0.85$
appear to better reflect China's rapid infrastructural development and
urbanization over the past decades, as the country moves from a largely agrarian
society to an industrial powerhouse.

In stark contrast, Japan exhibits remarkable consistency in the top-ranked
sectors, regardless of $\gamma$. The ``wholesale and retail trade; repair of
motor vehicles''~(26) sector consistently holds the top position, while sectors
such as ``construction''~(25) and ``motor vehicles, trailers and
semi-trailers''~(20) remain firmly rooted in the top ranks for both $\gamma$
values. This steadfastness showcases the well-established and mature industrial
and trade sectors of Japan, a hallmark of its developed economy.

The use of $\gamma = 0.85$, akin to the standard PR, aligns with prior
economic network research emphasizing the importance of network connections in
economic dynamics~\citep{page1998pagerank}. It appears to offer a balanced
interplay between value-added and network structures, ensuring that the economic
interactions among sectors are adequately represented. Given the notable shifts
observed in China's rankings between the two $\gamma$ values, it is evident that
while value-added considerations are crucial, the inter-sectoral connections and
interactions are still our focus. On the other hand, the stability of rankings
in Japan across both $\gamma$ values provides a testament to the resilience and
established nature of its economic sectors. We conjecture that the stability in
the top~5 sectors may be achieved after certain economic development stage when
using the extended PR method with value-added as auxiliary
information. The conjecture merits further investigation.

\section{Community structures}

\begin{figure}[tbp]
\centering
\includegraphics[width=\textwidth]{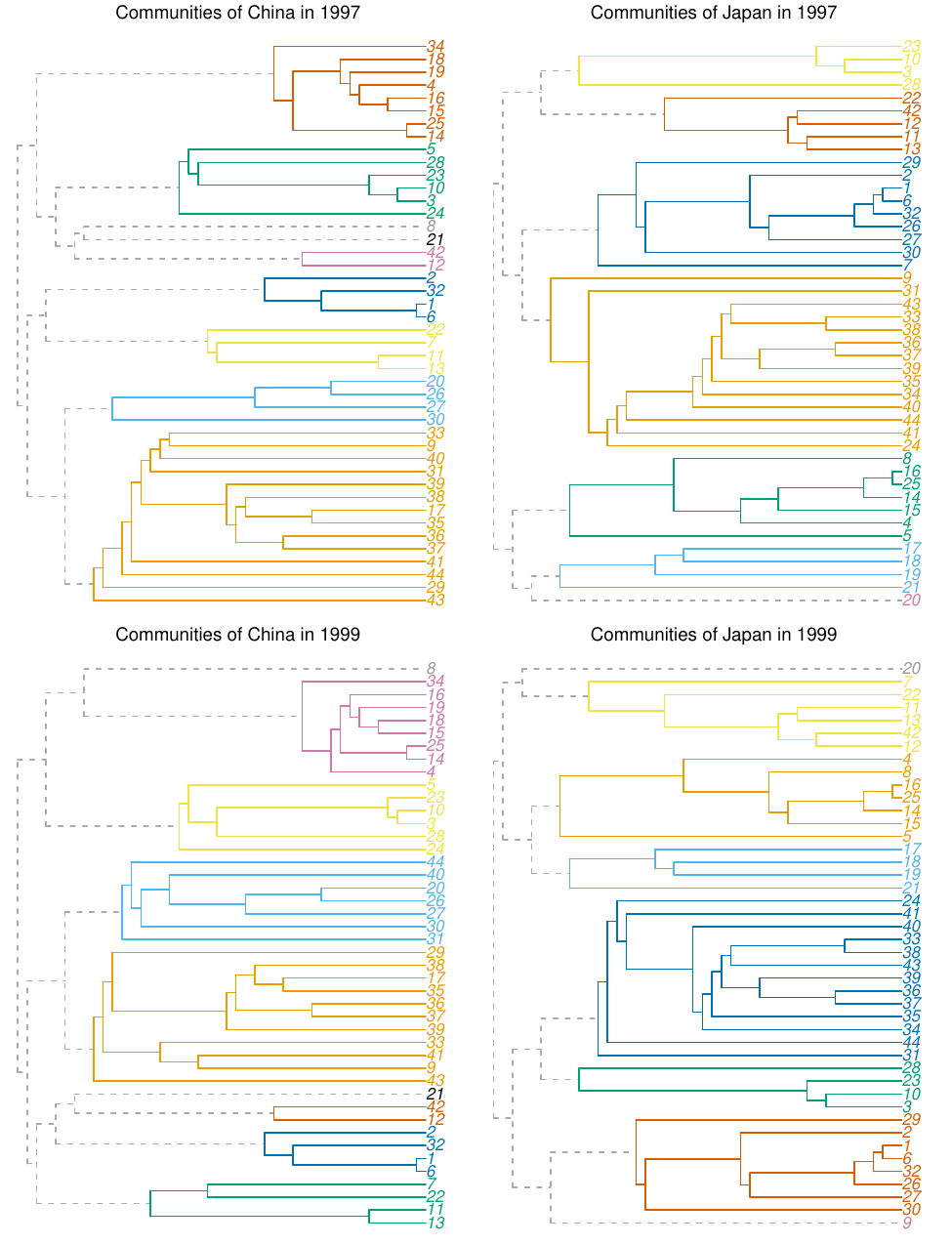}
\caption{Community detection of China and Japan in 1997 and 1999.}
\label{fig:supp1}
\end{figure}

\begin{figure}[tbp]
\centering
\includegraphics[width=\textwidth]{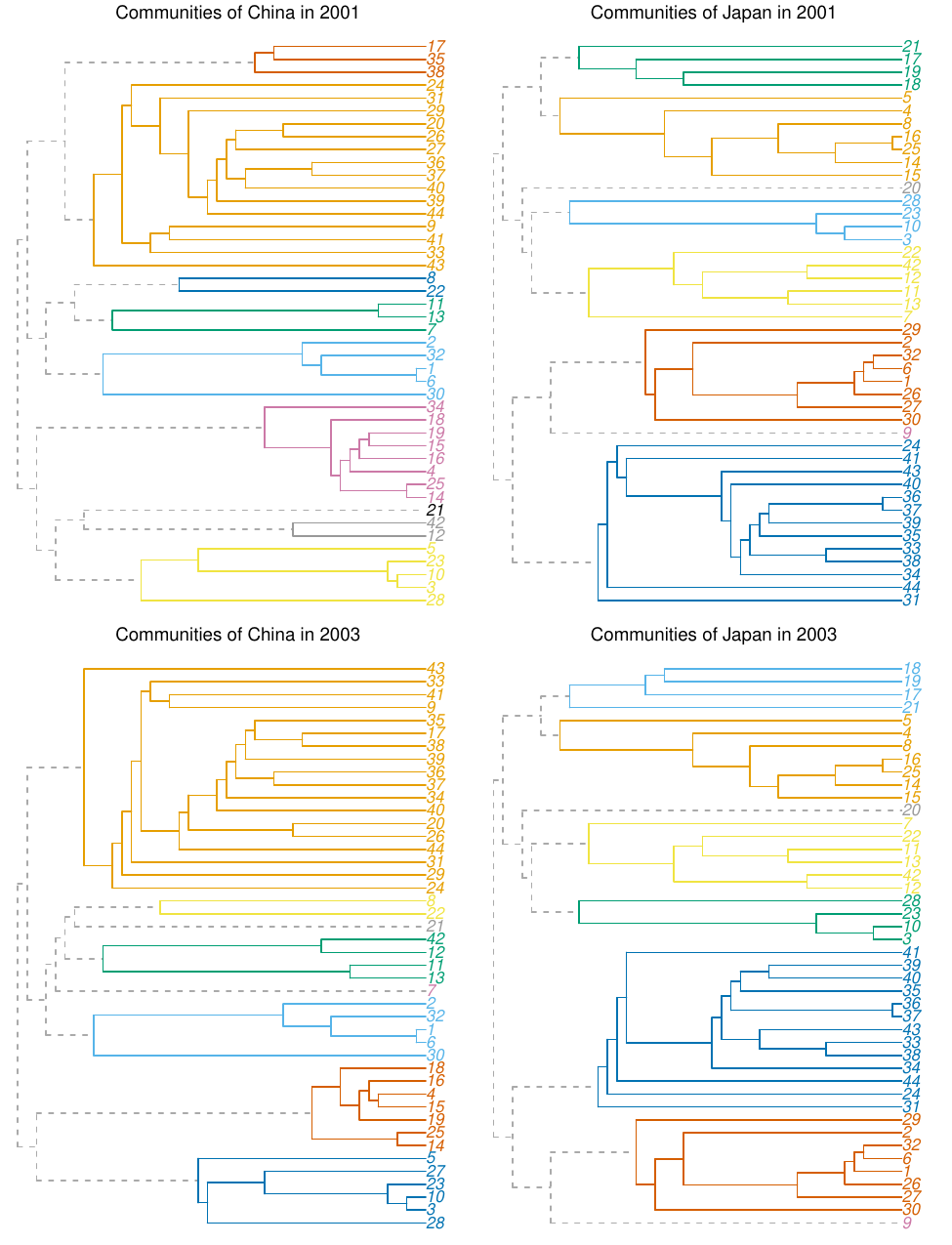}
\caption{Community detection of China and Japan in 2001 and 2003.}
\label{fig:supp2}
\end{figure}

\begin{figure}[tbp]
\centering
\includegraphics[width=\textwidth]{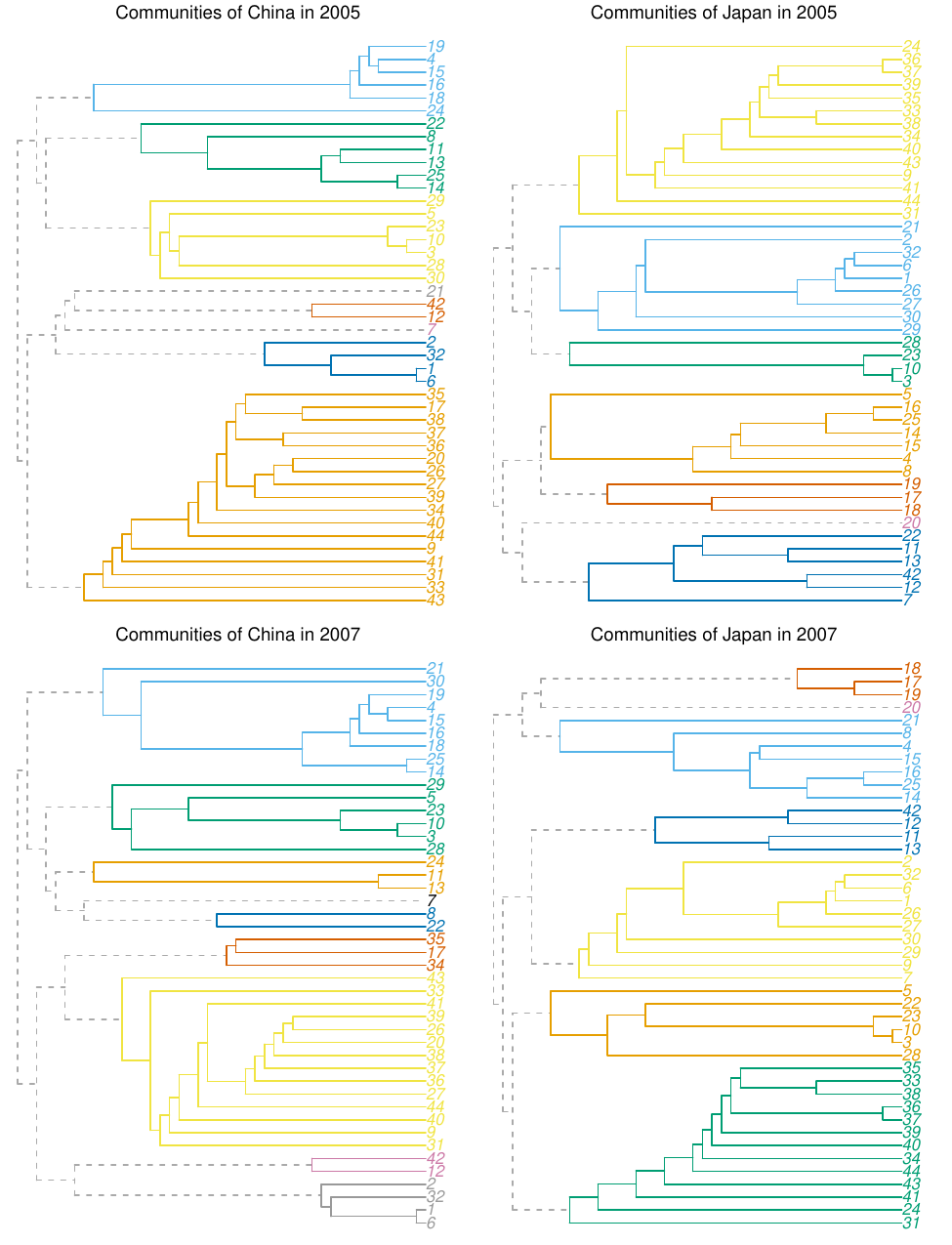}
\caption{Community detection of China and Japan in 2005 and 2007.}
\label{fig:supp3}
\end{figure}

\begin{figure}[tbp]
\centering
\includegraphics[width=\textwidth]{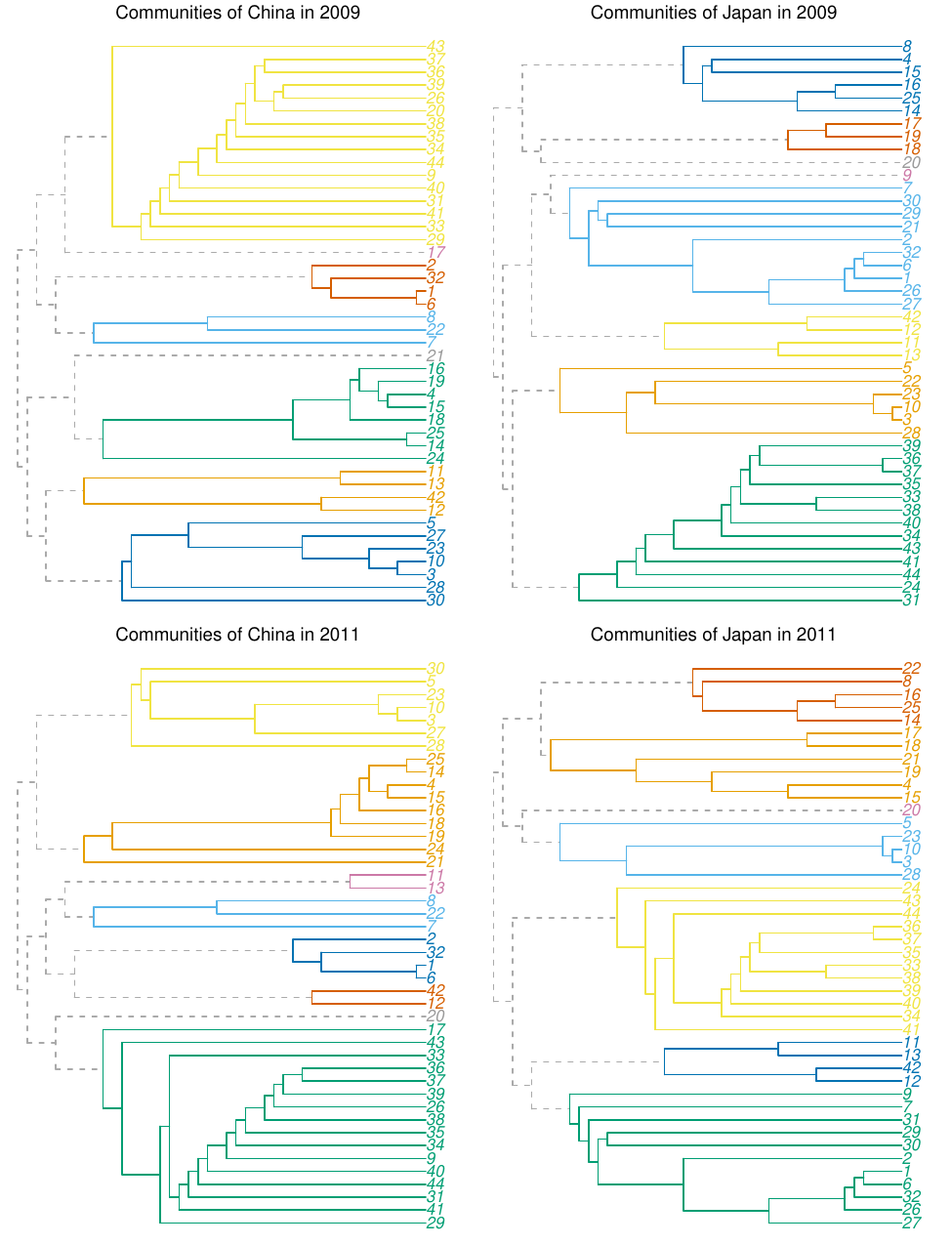}
\caption{Community detection of China and Japan in 2009 and 2011.}
\label{fig:supp4}
\end{figure}

\begin{figure}[tbp]
\centering
\includegraphics[width=\textwidth]{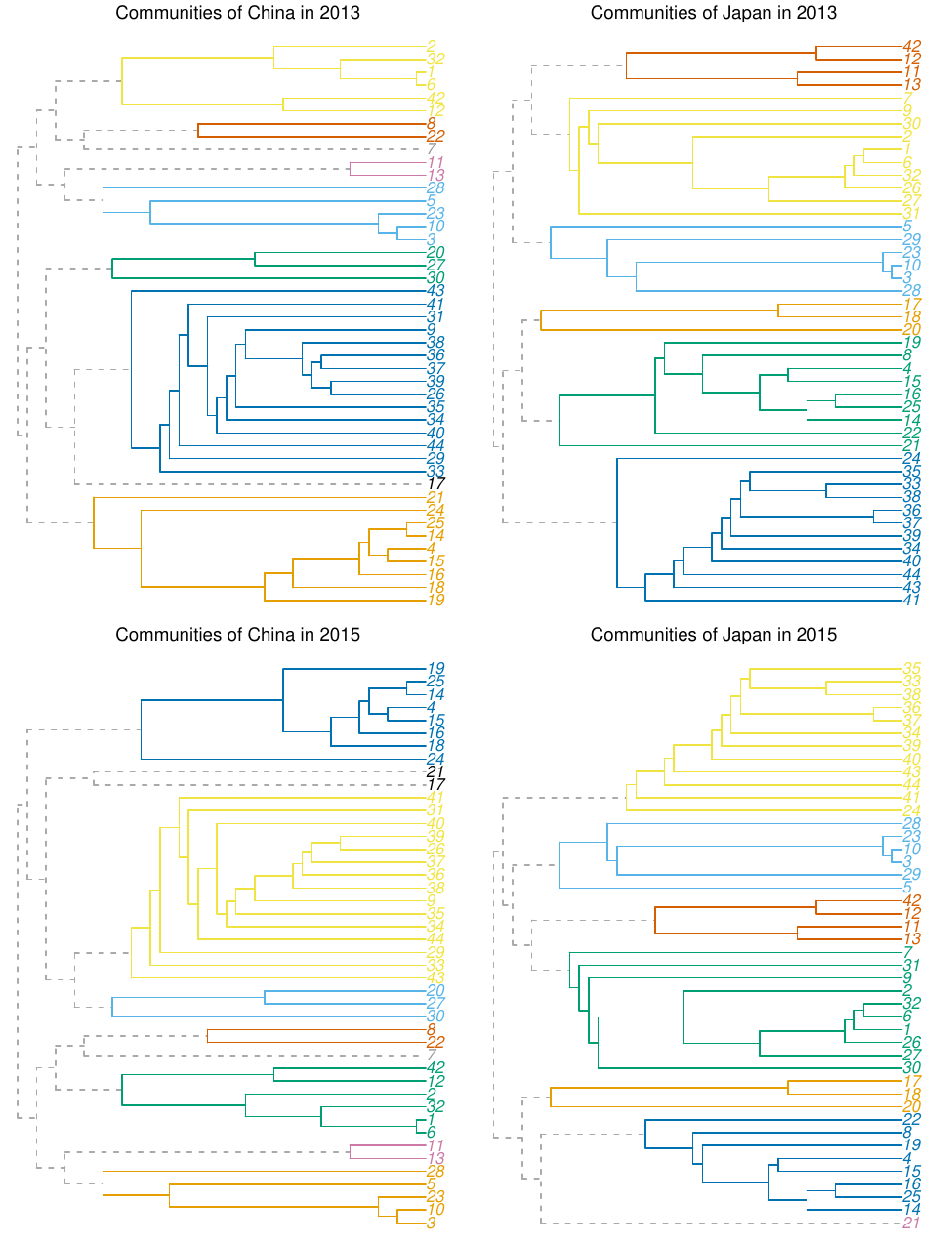}
\caption{Community detection of China and Japan in 2013 and 2015.}
\label{fig:supp5}
\end{figure}

Figures~\ref{fig:supp1},~\ref{fig:supp2},~\ref{fig:supp3},~\ref{fig:supp4} 
and~\ref{fig:supp5} provide the detailed information about the community 
structures of China and Japan for the years 1997, 1999, 2001, 2003, 2005, 2007, 
2009, 2011, 2013, 2015 and 2017.

\begin{figure}[tbp]
\centering
\includegraphics[width=\textwidth]{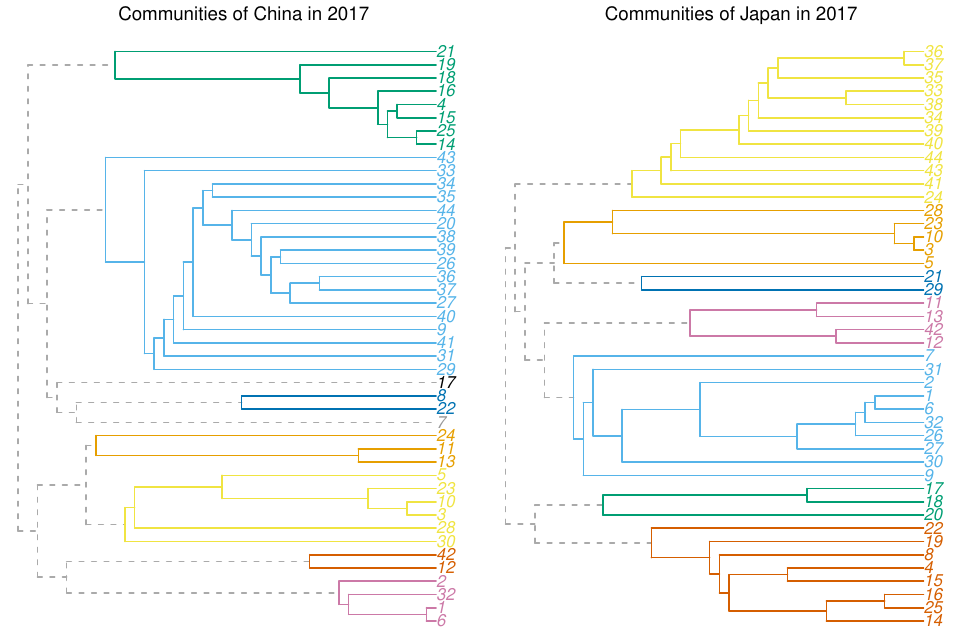}
\caption{Community detection of China and Japan in 2017.}
\label{fig:supp6}
\end{figure}

\bibliographystyle{chicago}
\bibliography{cjiot}